\documentstyle[aps]{revtex}

\begin{document}
\draft
%\preprint{Preprint }
\title{Comment on "Scaling of the quasiparticle  spectrum
for $d$-wave superconductors" }
\author{G.E. Volovik $^{(1,2)}$ and N.B. Kopnin $^{(2,3)}$}
\address{$^{(1)}$ Low Temperature Laboratory,  Helsinki
University of Technology, 02150 Espoo, Finland, \\
$^{(2)}$ Landau Institute for Theoretical Physics, 
117334 Moscow,
Russia,\\
$^{(3)}$ Laboratoire de Physique des Solides,
Universit\'e Paris-Sud, B{\^a}t 510, 91405 Orsay, France }

\date{\today}
\maketitle

%\eject
\narrowtext
\twocolumn

In a recent Letter Simon and Lee (SL) \cite{SimonLee}
suggested a scaling law for thermodynamic and kinetic
properties of superconductors with lines of gap
nodes. For example, the heat capacity as a function of
temperature  and magnetic field for $T\ll T_c$ and $H\ll
H_{c2}$ is
\begin{equation}
C(T,H)=aT^2 G(x)~ , ~x=\alpha T/H^{1\over 2}~,
\label{1}
\end{equation}
where  $G(x)$ is a dimensionless function of the
dimensionless parameter $x$. Eq. (\ref{1}) is
 in agreement with
our calculations
\cite{KopninVolovik,d-waveVortex,VortexEntropy} if
 the scaling
parameter is $x_{KV}= (H_{c2}/H)^{1/2}(T/T_c)$ 
(apart from  a
logarithmic factor). 
Indeed, according to \cite{KopninVolovik},
the function $G(x) \sim 1 +(1/x^2)$ for large values of $x$;
the first term is the bulk heat capacity $C(T,H)\propto
T^2$, while the second term results in a 
temperature-independent vortex contribution
$C(T,H)\propto H$. For small $x$,  the function
 $G(x) \sim 1/x$
which gives \cite{d-waveVortex,VortexEntropy}
$C(T,H)\propto T\sqrt{H}$. 
The crossover value between these
two regimes is $x_{KV} \sim 1$.
However, SL have obtained the crossover parameter
$x_{SL} \sim (H_{c2}/H)^{1\over 2}(T/T_c)\sqrt{E_F/T_c}$.
The difference between our $x$ and that obtained by SL is
thus by the large factor $\sqrt{E_F/T_c}$. We discuss the
origin of this disagreement.

Let us introduce the anisotropic Fermi 
momentum $p_F(\theta)$
which depends on the angle $\theta$ in the $a-b$ plane.
If $T \ll T_c$ the quasiparticles which are close to the
gap node, $\theta\ll 1$, are important. Their  spectrum is
\begin{equation}
E({\bf p})=\sqrt{v_F^2(p-p_F(\theta))^2
+(\Delta^\prime)^2\theta^2} ,
\label{2}
\end{equation}
where $v_F$ is Fermi velocity and $\Delta^\prime$ is the
the angular derivative of the gap,
$\Delta(\theta)\approx
\theta \Delta^\prime$, both are in a vicinity of the node.
Eq. (\ref{2}) was the starting point in
\cite{KopninVolovik,VortexEntropy}.

In contrast, SL used the linearized spectrum
\begin{equation}
E({\bf p})=\sqrt{c_\parallel^2\delta p_x^2 +c_\perp^2p_y^2} ,
\label{3}
\end{equation}
where ${\bf p}=p_y \hat y +
(p_F+\delta p_x)
\hat x$,  $c_\parallel=v_F$ and
$c_\perp=\Delta^\prime/p_F$. This is justified when the
nonlinear contributions to
\[
\epsilon ({\bf p})-E_F=p_x^2/2m_x+p_y^2/2m_y- E_F\approx
v_F\delta p_x+p_y^2/2m_y
\]
can be neglected, i.e., when $p_y^2/2m_y \ll c_\perp p_y$
where $p_y=p_F\theta $. This requires much stronger
restrictions both on the angle,
$\theta\ll T_c/E_F$, and on the energy and
temperature: $T \ll T_c^2/E_F$.

At the first glance, one might
expect that the temperature of order  of $T_c^2/E_F$
marks the boundary between our scaling and that by SL.
However, this is not the case. Our quasiclassical approach
is valid down to the temperature at which a discreteness of
fermion bound states in the vortex background becomes
important. For
$s$-wave superconductors, the interlevel spacing of core
fermions is of order of $ T_c^2/E_F$
\cite{Caroli1964}, thus the quantum limit is reached at
$T \sim  T_c^2/E_F$. In
$d$-wave superconductors, in a vicinity of the gap nodes, the
interlevel distance is smaller; it is determined by
a large dimension of the wave function which, for low
energies,  is limited by  the intervortex distance $R$
\cite{KopninVolovik,d-waveVortex}.  Thus the discreteness of
the levels becomes important at lower temperatures,
$T \sim (T_c^2/E_F) (\xi/R)  \sim (T_c^2/E_F)
\sqrt{B/B_{c2}}$.

Therefore, one expects two changes of
the regime with the crossover parameters as follows: (1) At
$x_{KV} \sim 1$ (i.e., at $
(H_{c2}/H)^{1\over 2}(T/T_c) \sim 1$)  the single-vortex
contribution to the thermodynamic quantitity is comparable
with the bulk contribution per one vortex.
(2) At $x_{KV} \sim
T_c/E_F$ (ie at $  (H_{c2}/H)^{1\over 2}(T/T_c) (E_F/T_c)
\sim 1$) the quasiclassical regime changes to the quantum
one. However, there is no change in the regime at the SL
scale, i.e., at $x_{SL} \sim 1$  ($(H_{c2}/H)^{1\over
2}(T/T_c)\sqrt{E_F/T_c} \sim 1 $). This is
because the high anisotropy of the conical spectrum in
Eq.(3) was not taken properly into account in
\cite{SimonLee}: the rescaling of coordinates
to get an isotropic spectrum of fermions with the average
"speed  of light"
$c=\sqrt{c_\parallel c_\perp}$ \cite{SimonLee} leads to a
high deformation of the potential well produced by
vortices.

\end{document}